# Damage-tolerant oxides by imprint of an ultra-high dislocation density

Oliver Preuß[1]*, Enrico Bruder[2], Jiawen Zhang[3], Wenjun Lu[3], Jürgen Rödel[1], Xufei Fang[1,4]*

[1]Division Nonmetallic-Inorganic Materials, Department of Materials and Earth Sciences, Technical University of Darmstadt, Peter-Grünberg-Str. 2, 64287 Darmstadt, Germany

[2]Division Physical Metallurgy, Department of Materials and Earth Sciences, Technical University of Darmstadt, Peter-Grünberg-Str. 2, 64287 Darmstadt, Germany

[3]Department of Mechanical and Energy Engineering, Southern University of Science and Technology, 1088 Xueyuan Avenue, Shenzhen 518055, P.R. China

[4]Current address: Department of Applied Materials, Karlsruhe Institute of Technology, Kaiserstr. 12, 76131, Karlsruhe, Germany

*Corresponding authors:

Oliver Preuß: preuss@ceramics.tu-darmstadt.de

Xufei Fang: xufei.fang@kit.edu

**Abstract:**

Dislocations in ductile ceramics offer potential for robust mechanical performance while unlocking versatile functional properties. Previous studies have been limited by small volumes with dislocations and/or low dislocation densities in ceramics. Here, we use Brinell ball scratching to create crack-free, large plastic zones, offering a simple and effective method for dislocation engineering at room temperature. Using MgO, we tailor high dislocation densities up to $\sim10^{15}$ m$^{-2}$. We characterize the plastic zones by chemical etching, electron channelling contrast imaging, and scanning transmission electron microscopy, and further demonstrate that crack initiation and propagation in the plastic zones with high-density dislocations can be completely suppressed. The residual stresses in the plastic zones were analysed using high-resolution electron backscatter diffraction. With the residual stress being subsequently relieved via thermal annealing while retaining the high-density dislocations, we observe the cracks are no longer completely suppressed, but the pure toughening effect of the dislocations remains evident.

**Graphical Abstract:**

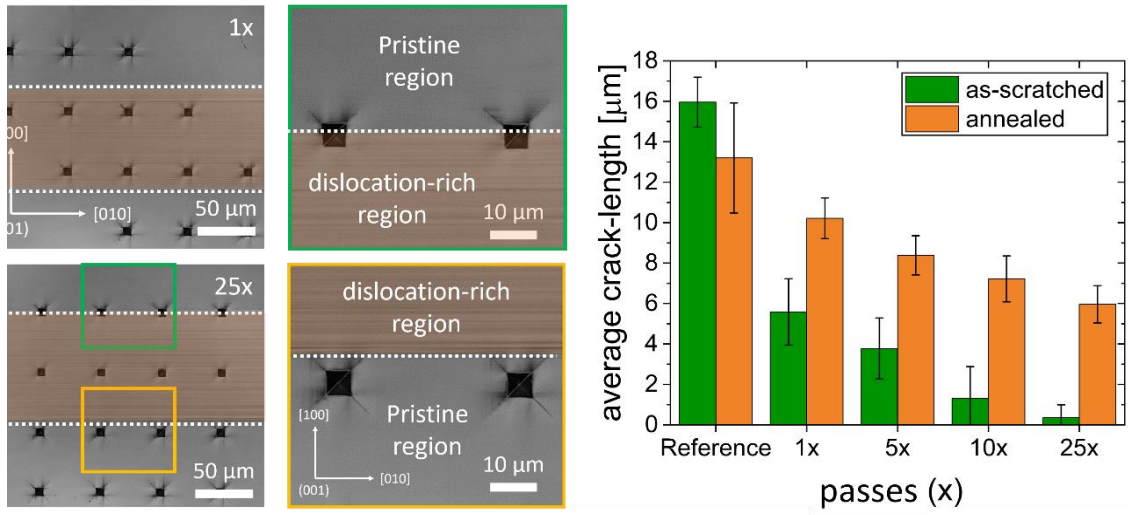

# 1. Introduction

The susceptibility of ceramics to flaws is the main limiting factor for their structural applications and their use as functional materials [1,2]. In typical brittle ceramic materials, defects such as pores or microcracks can lead to the failure of the whole component. These defects may be generated in the material during processing and machining. The propagation of a pre-existing crack in brittle solids has been well described scientifically, with the stress intensity factor ($K$-criterion) and the energy release rate approaches ($G$-criterion) being established for decades. Several toughening mechanisms have been developed for ceramics, which can be categorized into crack bridging, process zone mechanisms, and crack deflection [1]. Typically, the intrinsic fracture toughness of engineering ceramics is in the range of 0.5 – 3 MPa $\sqrt{m}$ [3], while fracture toughness of more than 10 MPa $\sqrt{m}$ can be reached via toughening mechanisms, e.g., by phase transformation toughening in partially stabilized zirconia [4].

Apart from these "traditional" synthesis-based methods, there has been a growing interest in dislocations in ceramics during the last years [5], mainly driven by the promising functional properties enabled by the charged dislocation core with ionic bonding (e.g., as electrical conductivity can be enhanced [6] and thermal conductivity can be lowered [7] for thermoelectric materials). For such dislocation-engineered ductile ceramics the dislocation behaviour at a crack tip is of eminent concern. For a better understanding of dislocation-based toughening in ceramics, there are four phenomena relevant and partly interfere with each other:

1. Shielding / anti-shielding: The presence of a dislocation in the vicinity of a crack-tip alters the apparent stress intensity based on its relative position and its Burgers vector. The dislocations that decrease the stress intensity are coined as the "shielding-type", while those that increase it are "anti-shielding-type" [8]. If a dislocation structure of mostly shielding-type dislocations can be achieved, a toughening effect can be observed, as was demonstrated in NaCl by Narita et al., where the fracture toughness was increased from 0.23 MPa $\sqrt{m}$ to 0.35 MPa $\sqrt{m}$ [9]. If the dislocation structure has a total Burgers vector of zero, there is no toughening by shielding [8,10].

2. Crack tip dislocation emission: A dislocation will be emitted if the stress level at a specific point in front of the crack-tip reaches the critical resolved shear stress of said dislocation. This specific point is half the dislocation core width from the crack-tip on the glide plane intersecting the crack front. The prerequisite is that the critical resolved shear stress is reached before the fracture stress, otherwise the crack will propagate instead of emitting dislocations. Emitted dislocations are always of the shielding-type [10,11].

3. Crack tip plasticity: Even if the yield criterion is not fulfilled for the bulk sample under loading, the stress in front of the crack tip can exceed the critical resolved shear stress due to local stress concentration. This means that dislocations inside this plastic zone become mobile and will move according to the applied stress. While the dislocations are gliding, mechanical work is done by the crack tip to overcome the lattice friction, which leads to toughening regardless of the Burgers vector [12, 13].

4. Blunting: In brittle materials, the crack-tip is considered to be atomically sharp[14]. If enough slip systems are available and crack tip dislocation emission is activated, the plastic deformation in front of the crack tip will counteract the crack propagation by blunting the tip. This can be understood on a dislocation level by the surface steps/ledges left behind on the fracture surface, when a shielding-type dislocation is emitted from the crack-tip. Crack tip blunting can effectively lower the stress intensity factor. The larger the tip radius, the lower the stress intensity factor [10, 15, 16].

Many proofs-of-concept have demonstrated the possibility of dislocation-based toughening in ceramics. In 1979, Appel et al. discovered a high dislocation density in front of a crack tip in MgO, similar to the plastic zone in front of a metal, using high voltage transmission electron microscopy (HVTEM). The effective stress around the crack-tip was verified as only slightly higher than in the bulk because of the present dislocations, resembling the shielding effect [17]. Moon and Saka demonstrated an increase in fracture toughness by a factor of ~2 in Yttrium-Aluminium-Garnet (YAG) using an annealing step to restructure the dislocations in sub-boundaries, which supposedly act as sources for the nucleation of more shielding-type dislocations [18]. Using a pre-engineering approach, where a high dislocation density is introduced mechanically prior to cracking the sample, Porz et al. [19] reached a surface dislocation density of ~$10^{14}$ m$^{-2}$ into single crystal $SrTiO_3$ (contrasting ~$10^{11}$ m$^{-2}$ on the pristine sample) and thereby increased the crack tip toughness by a factor of ~2. Using cyclic Brinell indentation to achieve a dislocation density 2-3 orders of magnitude higher than in the pristine single crystal, Preuß et al. [20] were able to increase the fracture toughness of $KNbO_3$ by a factor of up to 2.8. In these two cases, the dislocation density is assumed to be of random Burgers vectors and has no shielding effect. The toughening may therefore be caused by crack tip plasticity alone.

In all these cases using dislocations for toughening ceramics, the obtained fracture toughness is below ~1 MPa √m, which is still far from the values for FCC-metals that are well above 20 MPa √m [21]. Considering that pre-engineered dislocations in ceramics display a yet insufficiently high toughening effect for arresting *crack propagation*, here in this work, we take one step back and use dislocations to improve the resistance against *crack initiation*, namely, the damage tolerance. In contrast to the extensive studies on toughening mechanisms in ceramics towards suppressing crack propagation, crack initiation in ceramics has been less investigated. One of the better-evaluated fields is the crack initiation from grain boundaries in polycrystals by elastic anisotropy of the grains. Tvergaard and Hutchinson developed an analytical model for this in 1988 [22]. Yousef et al. later showed experimentally that this crack initiation by elastic anisotropy can be detected by temperature-dependent Young's modulus and thermal expansion measurements [23].

The prevention of crack initiation is a well-known concern in cyclically loaded metals, where fatigue limits the lifetime of components. A fatigue crack often starts from the surface, where the highest stresses (e.g. in bending) and stress concentration (notch effects from design features or surface roughness) can be found [24]. Surface hardening processes (e.g. shot-peening [25], laser shock peening [26], and deep rolling [27], to only name the mechanical ones) have been demonstrated to effectively protect the surface from crack initiation and/or crack propagation. The underlying mechanism relies on creating an extremely high dislocation density to work-harden the surface and forming compressive residual stresses in the range of GPa. Surface hardening has been deployed to a limited number of ceramics with promising results. For instance, Pfeiffer and Fry demonstrated that the load-bearing capacity of $Si_3N_4$ can be increased by a factor of 4 using shot peening to bring in a compressive residual stress of 2 GPa [28]. An increase in fracture toughness by up to 42% was achieved in SiC using laser shock peening by Shukla et al. [29]. The compressive residual stress induced was 0.2 GPa. The same group recently achieved a 60% increase in $Si_3N_4$ with compressive residual stresses near 0.3 GPa [30]. Worth noting is that, the mechanisms for these findings have barely been rationalized in terms of dislocations (although "micro-plasticity" was assumed by Pfeiffer and Frey [28], yet no direct evidence of dislocations was provided). Further, the applied load is usually very high and concentrated locally, which already introduces damage by itself.

Recent years have witnessed a rising interest in using dislocations in ceramics for tuning the mechanical properties. In particular, high dislocation densities (>$10^{13}$ m$^{-2}$) have been made accessible on a scale of hundreds of micrometers or even centimeters without cracking at room temperature [31,32]. Here, we begin by asking the following questions: Could we prevent the initiation of a crack via dislocation plasticity, so that the crack propagation only sets in at a higher load? Can we make ceramics less sensitive to pre-existing flaws by introducing a high dislocation density?

In this work, we engineer dislocation densities of ~$10^{15}$ m$^{-2}$ into single-crystal MgO using the cyclic Brinell ball-scratching technique [32]. The resulting dislocation structure is characterized regarding its size, depth, and density using chemical etching, electron channelling contrast imaging (ECCI) and transmission electron microscopy (TEM). Cracks are then introduced into these dislocation-rich zones using Vickers indentation to demonstrate the increased damage tolerance as indicated by the quantified crack length. The influence of compressive residual stresses is discounted by repeating the experiment with an additional annealing step before the cracks are introduced. The reduction in residual stress is verified by high-resolution electron backscatter diffraction (HR-EBSD). In the end, we discuss the mechanisms for dislocation-enhanced crack suppression and propagation.

2. **Experimental procedure**

## 2.1. Material

MgO single crystals were purchased from Alineason Materials Technology GmbH (Frankfurt am Main, Germany). MgO is of rock salt structure, and the room temperature active slip system is ½<110>{110}. The lattice friction is reported to be 65 MPa for pure edge-type dislocations and 86 MPa for pure screw-type dislocations at room temperature [33], suggesting very good room-temperature dislocation mobility.

The samples were of dimension 10 x 10 x 1 mm$^3$ with (100) orientation on the 10x10 mm$^2$ surface, where all experiments were conducted. Polishing of this surface in optical-quality (no surface scratches visible under the optical microscope up to 100x magnification) was done by the manufacturer and used as-received. The surface roughness $R_a$, obtained by confocal optical microscopy, is 4-6 nm. The pre-existing dislocation density will be examined and presented in the following sections.

## 2.2. Dislocation engineering

A high dislocation density was achieved using the cyclic scratching technique [32], which is schematically depicted in **Fig. 1A**. It was carried out on a universal hardness tester (Finotest, Karl Frank GmbH, Weinheim, Germany) with a hardened steel ball indenter (2.5 mm in diameter). The load applied was 9.8 N (1 kgf). The motion of the stage needed to achieve scratching during applied load was realized by mounting a motorized, computer-controlled one-axis stage (Physik Instrumente GmbH, Karlsruhe, Germany) on the hardness tester. The sliding velocity was set to 0.5 mm/s. Silicon oil was used as a lubricant to reduce wear damage of the indenter and the sample. The parameters (ball size, load, speed) were optimized beforehand, to exclude the damage of cracking and surface damage, while maximizing the achieved dislocation density in a feasible time frame.

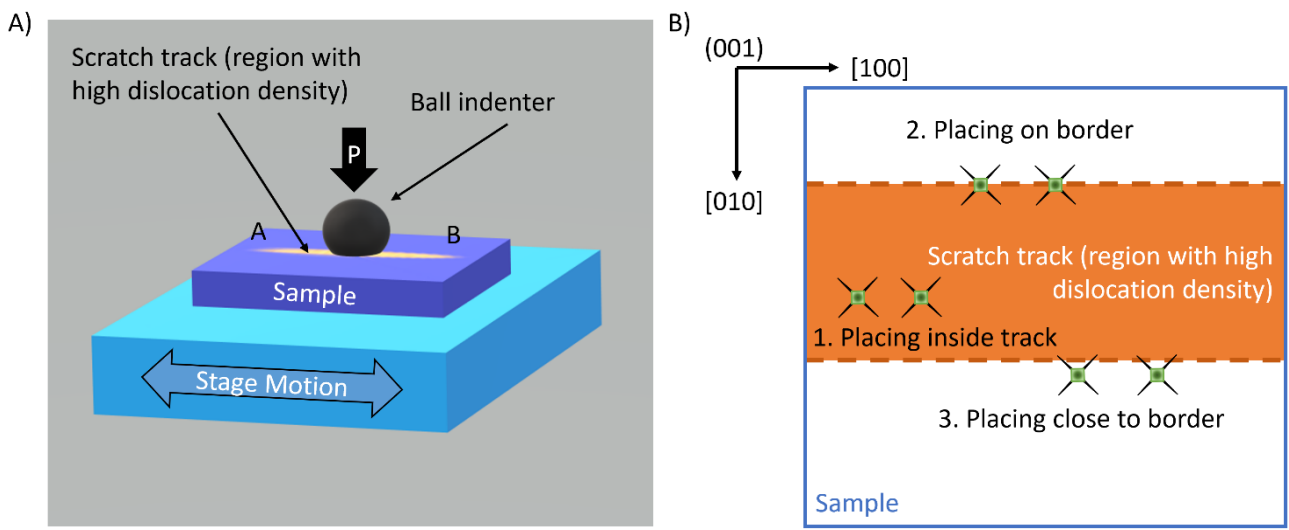

*Figure 1. A) Experimental setup for scratching the samples, B) different positions for placing Vickers indents relative to the resultant scratch track.*

By repeatedly scratching on the same track, different pass numbers can be achieved: Let A and B be the start and end point of the scratch track, then the movement AB would be termed one pass (1x), ABA two passes or one cycle (2x), ABAB three passes (3x), etc. In the current work, pass numbers with 1x, 5x, 10x, and 25x will be presented, as a saturation of dislocation density and mechanical properties was observed at higher pass numbers (50x, 100x).

The Brinell indenters (hardened steel ball, Habu Hauck Prüftechnik GmbH, Hochdorf-Assenheim, Germany) were checked regularly for changes in the surface roughness using a 3D confocal LASER scanning microscope (LEXT OLS 4000, Olympus IMS, Waltham, USA). Although it was found that the change was negligible even after 1000x, the indenter balls were rotated regularly to use fresh surfaces for scratching.

## 2.3. Characterization of Plastic Zones

### 1) Chemical etching

The etchant used for this work is 20 vol% sulfuric acid, with an etching time of 30 seconds at room temperature [34]. Etch pit patterns and densities were examined using a 3D confocal LASER scanning microscope (LEXT OLS 4000, Olympus IMS, Waltham, USA) with a Differential Interference Contrast (DIC) filter to highlight the surface topography changes.

One scratched sample was cut perpendicular to the scratch track to reveal its cross-section. After vibropolishing with 250 nm colloidal silica (OP-S) for 18 hours, the dislocation structure in the cross-sectional area (underneath the surface scratch track as in **Fig. 1A**) was again revealed by etching. Surfaces with high dislocation density were quantified with electron microscopy methods.

### 2) Electron channelling contrast imaging (ECCI)

Electron channelling contrast imaging (ECCI) was employed using a scanning electron microscope (MIRA3-XM, Tescan, Brno, Czech Republic) [35]. On {100}-planes in MgO, the dislocations appear as dots with tails and can be counted per area to estimate the dislocation density, even in areas where the values for density are not accessible to chemical etching ($>10^{14}$ $m^{-2}$).

### 3) Transmission electron microscopy (TEM)

TEM lamellas (~8×2×6 µm$^3$) were lifted out of 1x and 10x scratches parallel to the scratch direction with a dual-beam focused ion beam (FIB) (Helios Nanolab 600i, FEI, Hillsboro, USA). Scanning TEM / (STEM) was performed on these lamellas using a transmission electron microscope (FEI Talos F200X G2, Thermo Fisher Scientific, USA) at operating voltage of 200 kV. Annular dark field (ADF) imaging was done with probe semi-convergence angle of 10.5 mrad and an inner and outer semi-collection angle of 23-55 mrad.

**2.4. Mechanical testing: suppression of crack initiation and propagation**

To introduce controlled surface damage (crack formation), Vickers indentation offers an expedient and reproducible method. Typically, cracks are initiated from the four corners of the Vickers indenter and propagate into the material. The length of these cracks after unloading can be used to obtain semi-quantitative information on the fracture toughness [36, 37]. Experimental realization hinges on judicious choice of the Vickers load: If the load is too low, no cracks are initiated. If the load is too high, lateral cracks may penetrate the shallow plastic zone of about 100 µm underneath the surface and spallation may occur. In prior experiments, Vickers loads ranging from 0.098 N (10 gf) to 1.96 N (200 gf) were tested, with 0.98 N (100 gf) being the highest load without spallation under all test conditions (reference surface and 25x scratch track). Therefore, the load of 0.98 N was chosen for all following experiments.

Separation of crack initiation and crack propagation was achieved by placing the indents on or near the scratch track borders (dashed lines in **Fig. 1B**):

1. On the border, two of the four cracks are initiated outside the dislocation-rich zone, while two cracks are initiated inside the dislocation-rich zone.
2. Close to the border, all four cracks can be initiated outside the dislocation-rich zone, while two of them propagate into the dislocation-rich zone.

For better statistics in a semi-quantitative manner, 40 Vickers indents per pass number with a fixed indentation load of 0.98 N (100 gf) and a dwell time of 10 seconds were placed inside the scratch tracks as well as on the reference surface. The individual crack lengths were quantified (namely, 160 datapoints per condition, as each indent initiates four cracks). As Palmqvist cracks are expected in ductile ceramics [20, 36, 38], cracks are measured from Vickers indent corner to crack tip. In addition, to quantify the work-hardening in MgO, the same Vickers indents were also used to measure the Vickers hardness of the material with different dislocation densities.

**2.5. Residual stress quantification: before and after thermal annealing**

A whole section of the scratch track (150 µm x 350 µm) was covered by the HR-EBSD scan with a step width of 1 µm and an acceleration voltage of 15 kV. The data was evaluated using the software CrossCourt 4.5. By using the area outside the scratch track as stress-free reference, the residual stress was quantified. Furthermore, HR-EBSD can yield an estimate of the geometrically necessary dislocation (GND) density. The large step width, in combination of an assumption of a minimum strain energy, is expected to underestimate the GND-density due to incomplete information, but the data can be used as a lower bound estimate.

To release the residual stress, an annealing step was conducted at 1100 °C for 1h with a heating and cooling rate of 10 K/min. As validated by Salem et al. [38] on $SrTiO_3$, dislocations in ceramics can be stable using this annealing profile. After the annealing, HR-EBSD was repeated to quantify the stress after the annealing step.

## 3. Results and Analyses
### 3.1. Plastic zones induced by scratching
#### 3.1.1. Verification

The successful introduction of dislocations is indicated by the slip traces in **Fig. 2**. With increasing pass number, the slip traces get denser and finer, suggesting an increasing dislocation density by multiplication mechanisms [31]. As the density inside the scratch track increases, the gradient at the border gets more pronounced as well, as more and more slip traces protrude outside the direct contact zone. While the scratch track is roughly 100 µm wide, it is only 0.35 µm deep even after 50x (**Fig. 2F**). Note, that the imaging mode enhances the surface topography, making the scratch tracks appear deeper and the lines in scratching direction caused by surface asperities of the ball more prominent. All of these features are very straight, indicating no debris formed by wear, which would create more irregular patterns.

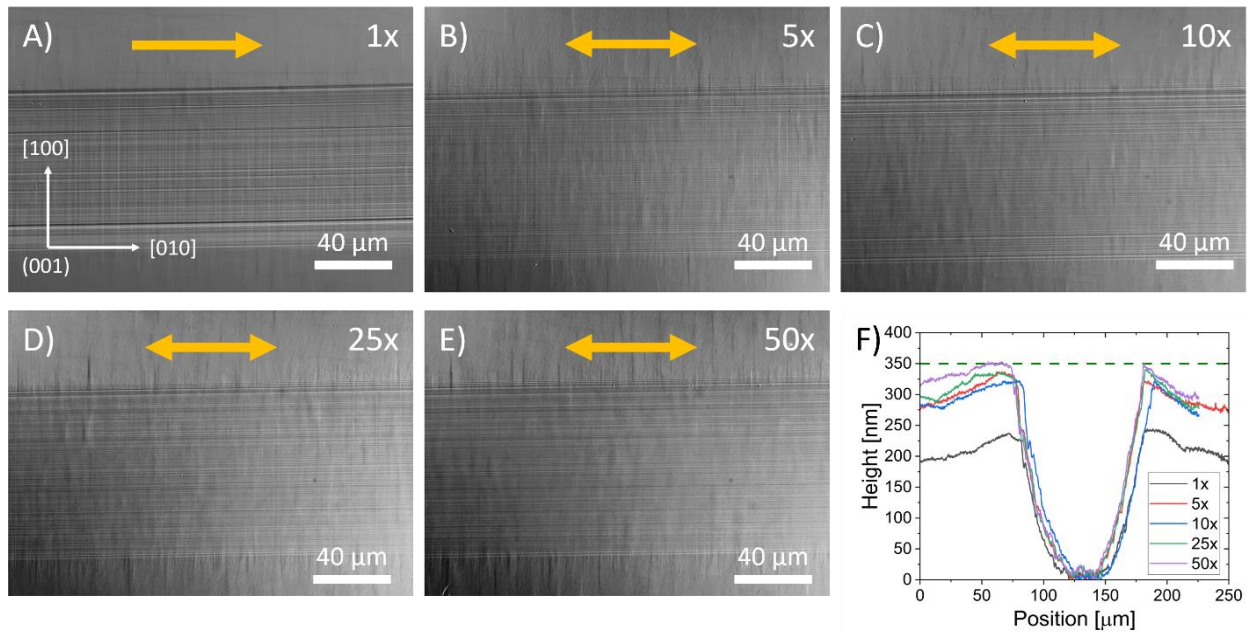

*Figure 2. Optical microscopy images of plastic zones induced by scratching with different pass number. A) 1x, B) 5x, C) 10x, D) 25x, E) 50x, F) depth profiles measured across the scratch track. Yellow arrows symbolize the scratching direction.*

#### 3.1.2. Estimation of dislocation density

For the reference sample, chemical etching with diluted sulphuric acid reveals a dislocation density (see **Fig. 3A**) of 7x10$^{11}$ m$^{-2}$. The same value for the reference surface is obtained using ECCI (**Fig. 3B**). The value for the pristine MgO dislocation density in literature is in the same range of 10$^{11}$ m$^{-2}$ to 10$^{12}$ m$^{-2}$ [17, 39].

Chemical etching is challenging for the scratch tracks, as the absolute number of dislocations is too high. It is also possible to estimate the dislocation density by the mean distance $\bar{d}$ of dislocations with the equation:

$$\rho_{DL} = \frac{1}{(\bar{d})^2} \tag{1}$$

where $\rho_{DL}$ is the dislocation density. Following this procedure, a mean distance of ~50 nm is obtained for 1x and ~25 nm for 25x from the ECCI images. This implies dislocation densities of ~4x10$^{14}$ m$^{-2}$ and ~2x10$^{15}$ m$^{-2}$, which marks an increase by a factor of ~600 (1 pass) and ~2800 (25 passes) as compared to the reference sample. The lighter and darker areas in the ECCI images are caused by slight misorientations of <0.1°.

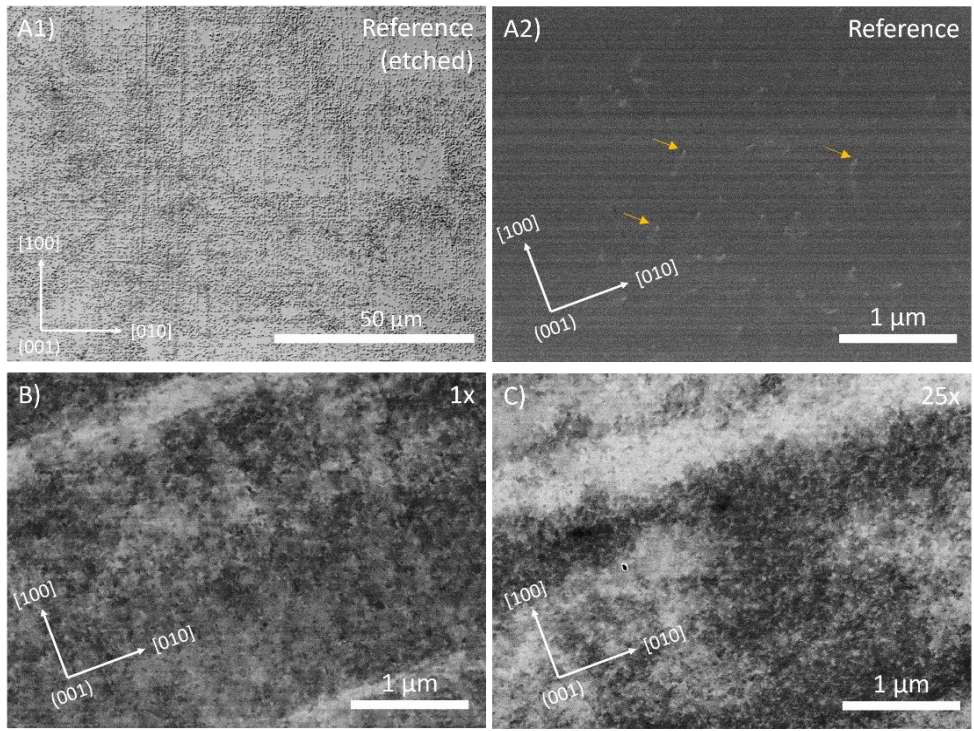

*Figure 3. Dislocation densities on deformed and undeformed surfaces. A1) pristine surface etched, A2) ECCI image of reference surface, B) ECCI image of 1x scratch track, C) ECCI image of 25x scratch track.*

Considering the cross-section (**Fig. 4**), the dense dislocation structure reaches ~150 μm deep into the material. The distribution of the dislocations becomes more uniform for higher pass numbers (>10x) down to ~50 μm.

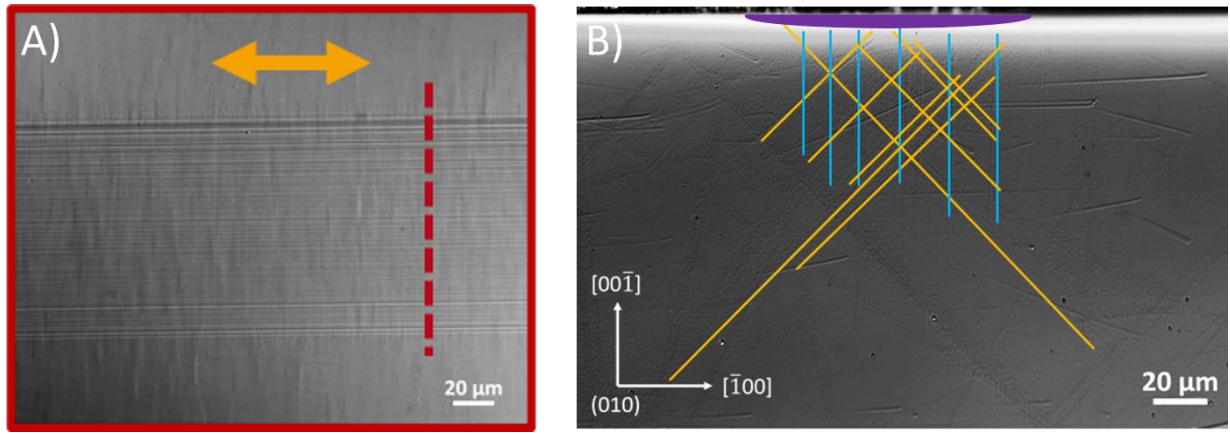

*Figure 4. Sample cross-section etched with 20 vol% sulfuric acid to reveal the dislocation structure. A) schematic of cross-section position, B) 10x scratch track with relative position of the scratch track (purple marker). The dislocation slip traces are marked with yellow lines (45° to surface) and blue lines (90° to surface) for better visualization.*

### 3.1.3. Dislocation structure in TEM

Comparing the TEM lamellae of the as-scratched state, a clear gradient of dislocation density can be observed (**Fig. 5**). With 10x scratching, the dislocations protrude further into the material than for 1x. The high density and the image quality don't allow for a quantitative analysis of the dislocation density. However, using a simple line-wise average greyscale analysis of the final image, the gradient into the sample can be fitted using an exponential function which yields an approximate correlation length for the gradient. This characteristic length is 0.4 µm for 1x and 2.4 µm for 10x, meaning that the dislocations travel ~6 times further into the material with higher pass numbers. Because of the inhomogeneous dislocation density (e.g. seen in **Fig. 3**), the characteristic length might vary at different positions within the scratch track.

The TEM lamellae in the annealed state (**Fig. 5B**), on the other hand, both exhibit cell-like structures of dislocation lines. This verifies the onset of recovery and polygonization (as temperature increases the enhanced mobility of dislocations enables cross-slip, allowing obtainment of an energetically more favourable configuration), a process well known from highly-deformed, dislocation-rich metals. This renders a more ordered dislocation structure after annealing.

The parallel 45° lines in the as-scratched 1x (**Fig. 5A1**) state further suggest that the indenter scratched over the surface sends out half loops in front of it on the {110} planes, conforming to the slip system in MgO. As the scratching was performed in a back-and-forth manner, the generated dislocations were forced to interact, which opens the possibility for multiplication and forest hardening. This mechanism would rationalize the "messy", noodle-like structure with a much higher dislocation density in the 10x case (**Fig. 5A2**).

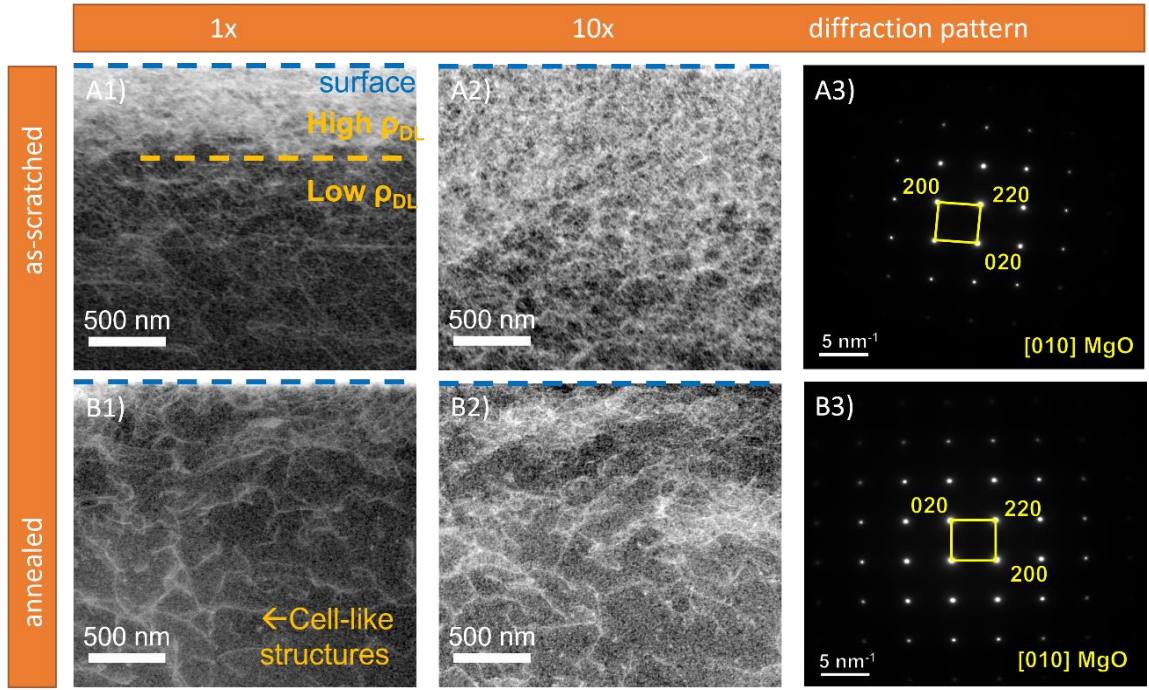

*Figure 5. ADF-STEM observation of dislocation structure on sub-micrometer scale along a 1x and 10x scratch track before (A) and after thermal annealing at 1100 °C for 1 h (B). The upper edge of each image represents the sample surface (blue dashed line).*

**3.2. Mechanical testing**

The Vickers hardness calculated from the size of the residual imprint increases with increasing dislocation density. A maximum enhancement in hardness of ~40% was achieved (**Fig. 6**). This demonstrates the work-hardening behaviour of MgO, which is well known in the literature [40]. The work-hardening in MgO is attributed to the formation of sessile junctions (Lomer-locks), resulting in severe forest hardening. Note that due to the crack formation in certain cases (reference sample and plastic zones with 1-pass scratching), the absolute values in the Vickers hardness shall be used with caution.

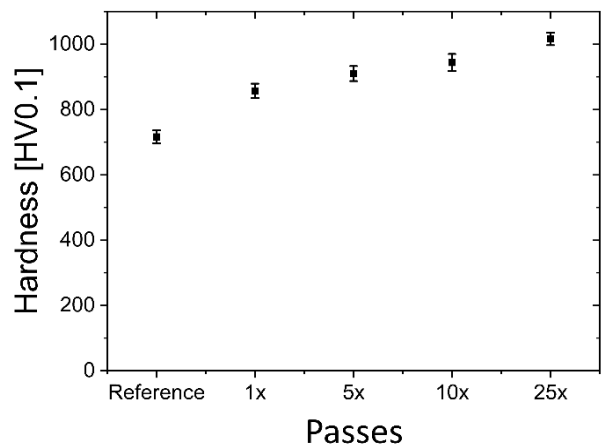

*Figure 6. Vickers Hardness of as-scratched samples as a function of number of passes.*

The length of the cracks introduced by Vickers indentation does not change much when comparing 1x and reference (see **Fig. 7A**). With 5x, the crack length decreases significantly, while with 10x and 25x cracking is avoided altogether. By placing Vickers indents directly on the border of a 25x scratch track (green rectangle in **Fig. 7B1**), crack initiation is proven completely suppressed inside where the high-density dislocations were present. If the Vickers indent is placed close to the border, the crack is initiated outside the scratch track (where the dislocation density is much lower) and propagation is stopped at the border (**Fig. 7B2**).

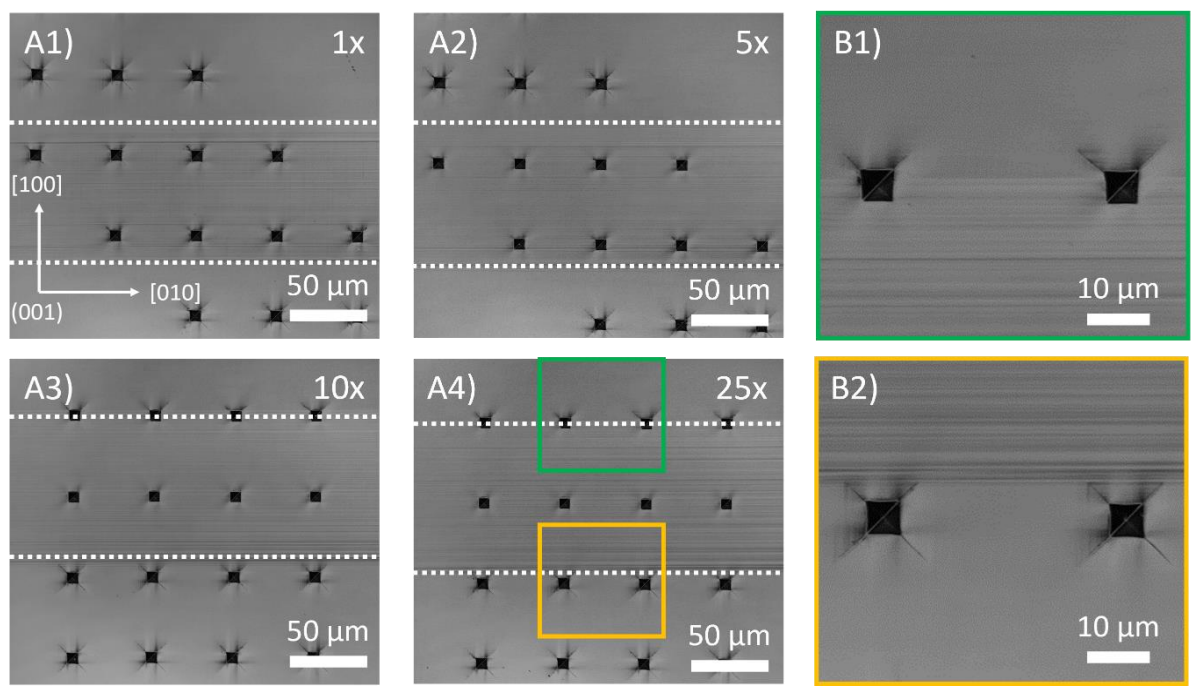

*Figure 7.A) Vickers arrays at scratch tracks with pass number of 1x, 5x, 10x, 25x (from A1 to A4). The white dashed lines mark the border of the scratch track. B) enlarged images of Vickers indents directly on (B1) and close to (B2), the border of the scratch track.*

A more quantitative analysis in **Fig. 8A** verifies crack shortening as a function of pass count, with the most significant change evidenced between the reference state and 1x scratching. Noteworthy is that ~45% of all the cracks are completely suppressed (crack length is zero) with 10x, and cumulative probability for complete crack suppression increases to ~70% with 25x. For very short cracks (i.e., <1-2 µm), the crack length itself approaches the error of measurement, which could lead to added scatter in the data for higher pass counts. As a high defect density eliminates the long tail at high crack length, the failure of ceramics appears more predictable, rendering a steeper slope of the cumulative probability plot (namely, less scatter). There is no such trend visible in the as-scratched state.

Another possible explanation is, that, after annealing, the reference state demonstrates irregular crack patterns, which results in high scatter. This represents the irregular crack growth of these indents: in most cases, more than 4 cracks per indent grew, the cracks initiated from edges in some cases, and their paths appear curved or deflected. The crack patterns inside the scratch tracks were regular with 4 cracks emanating from the indenter corners.

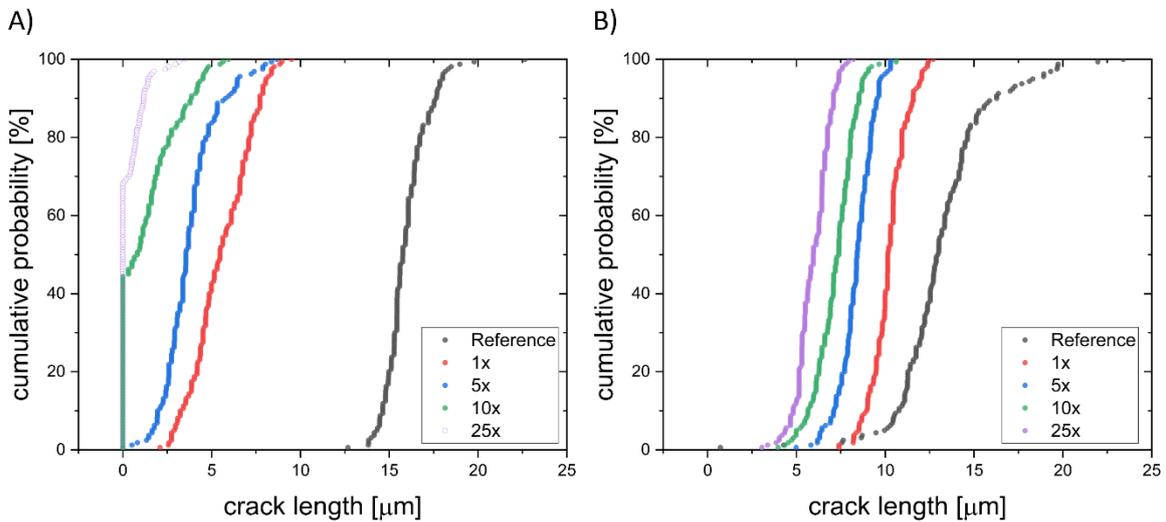

*Figure 8. Cumulative probability plot of Vickers crack lengths quantified from Vickers corner to crack tip. The Vickers indents were introduced into A) the as-scratched state, B) after annealing of the scratched sample.*

### 3.2.1. Residual stresses before and after annealing

By performing HR-EBSD before and after annealing of a scratched sample, the stress state relative to the surrounding reference area can be quantified. The shear components ($\tau_{12}$, $\tau_{13}$ and $\tau_{23}$) and the out-of-plane stress component ($\sigma_{33}$) are in the MPa range, while the in-plane principal stress components ($\sigma_{11}$ and $\sigma_{22}$) reach values of -1.53 GPa (1x) to -1.70 GPa (25x) as compared to the pristine surface averaged over the track (see **Fig. 9** where $\sigma_{11}$, $\sigma_{22}$ is similar, with the minus sign signifying compressive stress). These high compressive residual stresses act as closure stresses on the cracks introduced using Vickers indentation.

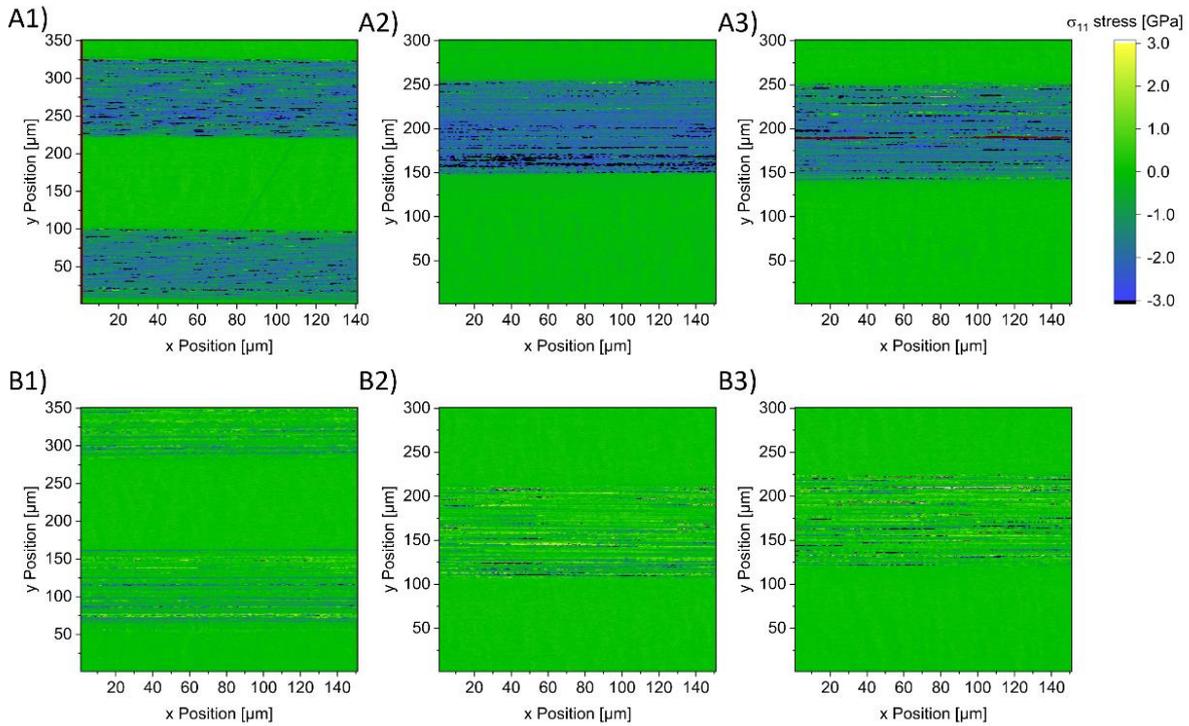

*Figure 9. HR-EBSD maps of $\sigma_{11}$ stress (lateral normal stress component) before and after annealing for scratch track sections with different pass numbers. A) as-scratched state: A1) 1x (top) and 5x (bottom), A2) 10x, A3) 25x. B) depicts the annealed state in an analogous manner.*

After annealing, $\sigma_{11}$ and $\sigma_{22}$ are lowered drastically to -0.26 GPa (1x) to -0.02 GPa (25x) on average. The remaining residual stress could be attributed to artifacts from the internal separation of elastic and plastic strain tensors by the image correlation software.

The HR-EBSD for the as-scratched sample verifies the increase in dislocation density (**Fig. 10**) seen in chemical etching and ECCI (**Fig. 3**). With higher pass number, the average density of geometrically necessary dislocations at the track increases from $3 \times 10^{13}$ m$^{-2}$ (1x) to $5 \times 10^{13}$ m$^{-2}$ (25x). It also confirms its inhomogeneous distribution in the scratch track: the dislocation density appears to be slightly higher towards the scratch track border than in the center. This is probably caused by the complex stress state under a Brinell ball indenter.

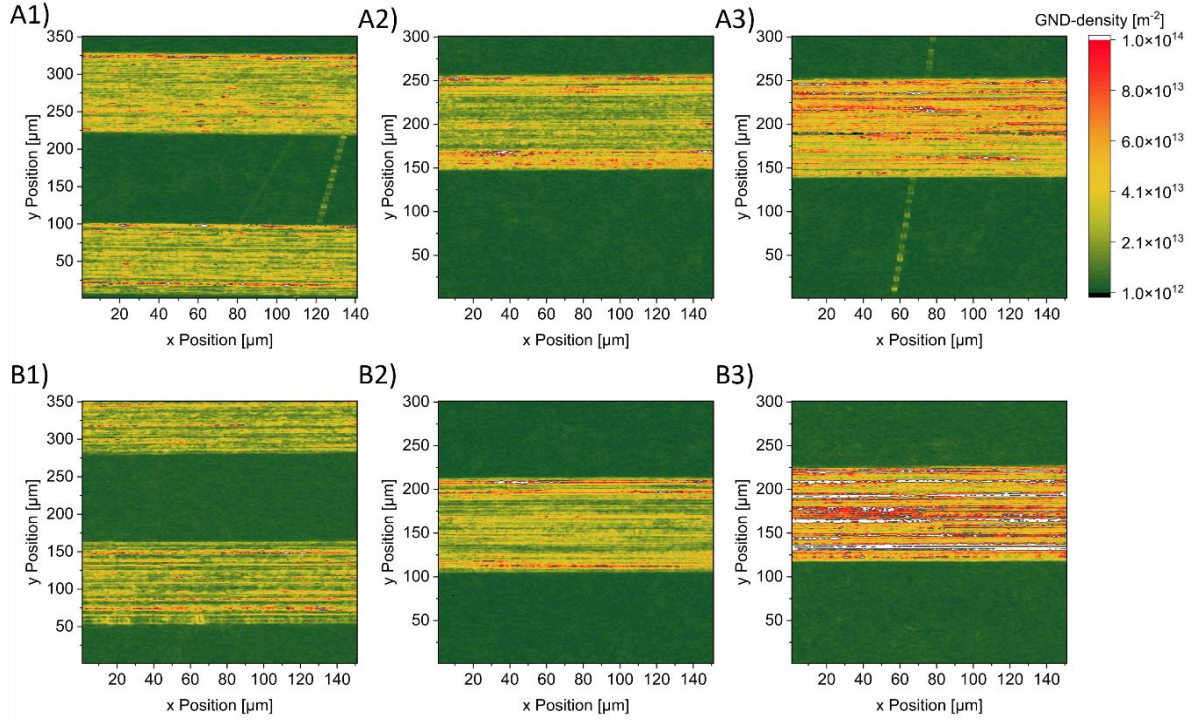

*Figure 10. GND density before and after annealing for scratch track sections with different pass number. A) as-scratched state: A1) 1x (top) and 5x (bottom), A2) 10x, A3) 25x. B) reveals the annealed state in an analogue manner.*

After annealing, no significant change in dislocation density is observed: the average density of dislocations at the track ranges from $3\times10^{13}$ m$^{-2}$ (1x) to $7\times10^{13}$ m$^{-2}$ (25x), comparable to the values before annealing. Therefore, the sample can be considered free of residual stress with retained high dislocation density.

## 4. Discussion

### 4.1. Dislocation structure

The dislocation densities by cyclic scratching of MgO are almost 2 orders of magnitude higher than reported for the well-investigated SrTiO$_3$ under similar deformation conditions (~ $10^{15}\ m^{-2}$ vs. $10^{13}\ m^{-2}$ [32]). This implies a higher ductility of MgO, which can be rationalised by its shorter Burgers vector ($|b| = \sqrt{2}a$ for perovskites, $|b| = \frac{\sqrt{2}}{2}a$ for NaCl-structure, where $a$ is the lattice parameter, assuming the <110>{110} slip system at room temperature). Furthermore, it is known from literature[41,42] that dislocations in SrTiO$_3$ can split into partials at room temperature, while no such behaviour is reported for MgO. Instead, room-temperature dislocations in MgO exhibit extended core structures [42]. Partial dislocations hinder cross-glide, confining the dislocation motion in SrTiO$_3$ to narrow slip bands of high dislocation density. In contrast, dislocations in MgO can leave their glide plane in case a pile-up occurs, resulting in a more homogeneous distribution. The mobility of dislocations in MgO is mostly governed by lattice friction and the presence of other defects [43,44]. This latter dependence is also showcased by the immense work-hardening, caused by Lomer-locks resulting from dislocation reactions.

From the loading conditions and the ECCI measurements, it can be assumed that the Brinell ball generates dislocation half-loops in front of it while sliding against the material. On the back-stroke, the new dislocation half-loops are forced to intersect with the old ones when expanding, resulting in dislocation reactions and multiplication events, increasing the dislocation density and the forest hardening. In the as-scratched state with high pass counts, the dislocation structure is highly chaotic with low mobility due to strong dislocation forest hardening.

During the annealing procedure, the mobility is increased [44] and the dislocations can rearrange in a low-energy configuration, which is indicated by the cell-like structures in the TEM images (**Fig. 5B**). This process is similar to recovery in severely deformed metals, including the formation of polygons. This means that the dislocation mobility is decreased even more, the dislocations are not as evenly distributed, and they can hardly contribute to toughening.

### 4.2. Influence of compressive residual stresses

The high compressive residual stresses observed directly after scratching (**Fig. 9A**) are clearly the major contributors to the significantly improved damage tolerance in addition to the high-density of dislocations. The residual stress stems from the near-plastic regions in the complex stress state during deformation, which release the high elastically stored energy once the Brinell ball has passed. This phenomenon can be leveraged in future structural material applications to achieve complete crack suppression [45, 46].

To discern the potential coupling of residual stresses and dislocations, annealing gives an indication towards the true contribution of dislocations. Annealing removes the residual stress whilst retaining most of the ultra-high-density dislocations, although they may be rearranged. TEM observation suggests that the dislocation density might be reduced by up to one order of magnitude during annealing (**Fig. 5**). However, the dislocation density in the scratched and annealed state remains ~100 times higher than in the pristine one. Under as-scratched conditions, the average crack length in 25x scratch tracks can be shortened by 98%, but only by 53% in the annealed state (see **Fig. 11**). The smaller effect in comparison might also be caused by the reduced mobility of dislocations due to polygonization. The rearrangement of the dislocations might well be the reason for the stress relief.

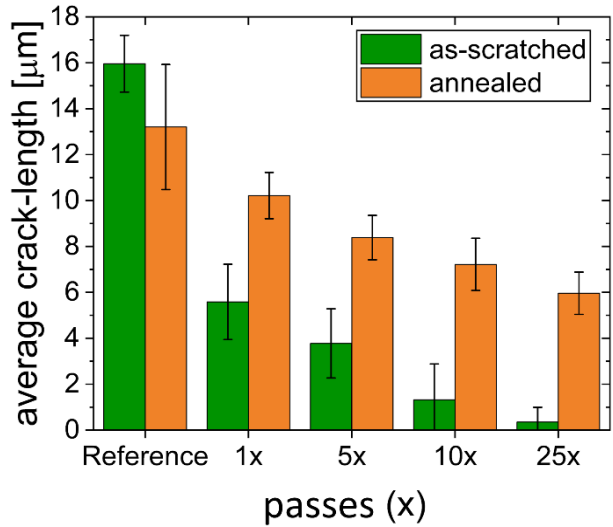

*Figure 11. Average crack lengths gained from 160 Vickers-induced cracks per condition.*

**4.3. Dislocation-based damage tolerance and toughening**

As mentioned in the introduction, other than the obvious dominance of the compressive residual stresses, several crack tip-dislocation interaction scenarios can achieve resistance against crack initiation or crack propagation. The dislocation structure in front of the cracks in this work can be assumed as a dislocation cloud of random Burgers vectors, which means the total shielding of the pre-engineered dislocations is zero [8]. However, once activated by the crack tip stress field, they can act as multiplication centres and sources for shielding-type dislocations. Because crack tips in MgO can already emit dislocations by themselves (at least in thin samples [11, 17] with slow crack propagation), the activation of such sources seems likely. This aided emission would be especially effective at medium dislocation densities, where the behaviour changes from nucleation-dominated to multiplication- and motion-dominated. A very high dislocation density, like in the 25x case, offers many nucleation sites but also decreases the dislocation mobility drastically due to work-hardening (**Fig. 6**), rendering the newly nucleated dislocations ineffective. The results do not seem to indicate such an optimal dislocation density, as the crack length decreases continuously with dislocation density in all cases. Worth noting is that, the crack tips observed in this work also exhibit no sign of blunting.

A random dislocation cloud contributes to crack tip plasticity as a dynamic effect during crack propagation, as shielding-type dislocations are repelled and anti-shielding-type dislocations are absorbed within the plastic zone where the stresses exceed the lattice friction. The amount of mechanical work $W_{mech}$ consumed by moving a section of dislocation of length $\partial l$ by a distance $\partial s$ can be assumed by:

$$\frac{\partial W_{mech}}{\partial l\, \partial s} = \tau \cdot |b| \approx \tau_{CRSS} \cdot |b| \tag{1}$$

where $\tau_{CRSS}$ is the critical resolved shear stress and $|b|$ is the length of the Burgers vector. For MgO, this results in 0.026 J/m² per dislocation, while the fracture energy should be approximately equal to $2\gamma \approx 13.7$ J/m², where $\gamma$ is the surface energy of the cleavage plane [47]. With this estimation, it requires hundreds of dislocations in the plastic zone to significantly impact the fracture behaviour. Following the expressions by Thomson [8], the plastic zone size should be approximately 10 µm by 2 µm in cross-section, meaning the dislocation density needed for toughening effects has to be at least $10^{13}$ m$^{-2}$. However, with this density (between pristine and 1x), work-hardening already plays a role (**Fig. 6**), which increases the average needed stress to move a dislocation. This in turn reduces the plastic zone size where sufficient stress is provided, so a higher density is needed which leads to more work-hardening. This paradox was addressed by Hartmaier and Gumbsch for BCC tungsten [48] and raises a new pertinent question: is there a dislocation density in MgO, where the pre-engineered dislocations act as sources so there is always at least one active source in the vicinity of the crack tip, but the mobility is not hindered too much so the nucleated dislocations can participate in crack tip plasticity? A comparison to a material with similar structure and plasticity but less work-hardening may help to verify or falsify these hypotheses. The polygonization in the annealing process lowers the mobility even more and reaching a toughening effect gets more difficult.

To summarize the possible mechanisms for dislocation-enhanced crack suppression, **Fig. 12** illustrates the scenarios for different dislocation densities schematically. With no or few dislocations present, the Vickers indent leads to typical brittle behaviour and Palmqvist cracks are formed (see **Fig. 12A**). With 1x, a high surface dislocation density is reached, while lower regions (beneath ~0.5 µm depth) show lower dislocation density (compare with **Fig. 5**). The result of lower pass numbers can be influenced by this sharp gradient, as the stressed volume of the Vickers indenter (several micrometers deep) intersects both the regions with higher and lower dislocation density along the depth (see **Fig. 12B**) and cracks can still be initiated lower down. In a 10x scratch track, the gradient in dislocation density is less steep and the stressed volume is completely filled with the ultra-high dislocation density coupled with high residual stress, with no cracking taking place (**Fig.12C**).

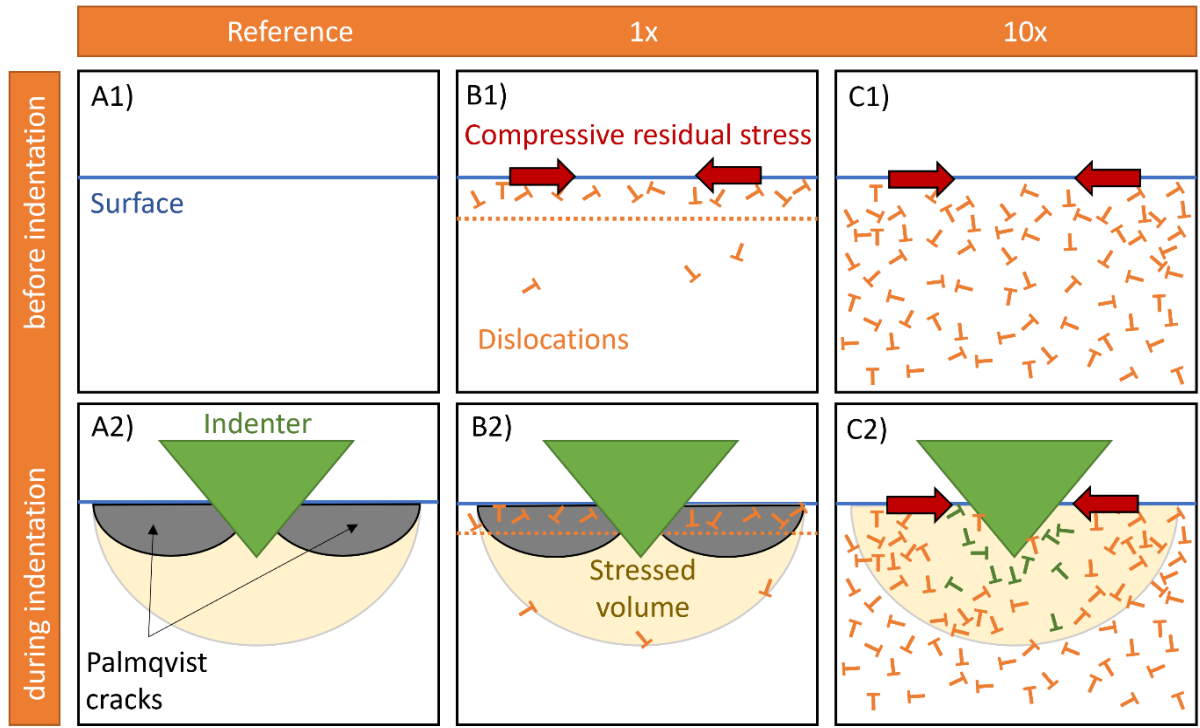

*Figure 12. Schematic of Vickers indentation with no (Reference, a), low (1x, b) or high (10x, c) pre-existing dislocation density.*

Note that the quantification of damage tolerance by crack lengths initiated by Vickers indentation does not afford differentiation between resistance to crack initiation and crack propagation, as both mechanisms occur simultaneously with this method. The Vickers indentation itself, combined with the present dislocations acting as sources, may even facilitate further dislocation multiplication and motion underneath the indenter to consume a portion of the input energy during the Vickers indentation process (see green dislocations in **Fig. 12C2**). The complete absence of cracks in the as-scratched state with 10x and 25x, however, clearly demonstrates the resistance against crack initiation facilitated by the engineering of ultra-high density of dislocations.

## 5. Conclusions

An ultra-high dislocation density has been introduced in single-crystal MgO at room temperature by cyclic scratching with a Brinell indenter. The resulting dislocation-rich zone has no limitation on length, a width of ~150 µm and a depth of ~150 µm. The dislocation density reaches up to $2 \cdot 10^{15}\ m^{-2}$ after 25 passes and can be tailored over three orders of magnitude by adjusting the number of passes. Crack initiation and propagation caused by 0.98 N (100 gf) Vickers indents can be completely suppressed in the as-scratched state. With the synergistic high dislocation density and high compressive residual stress (up to 2 GPa), the dislocation-engineered samples exhibit excellent damage tolerance, namely suppression of crack formation. By releasing the high compressive residual stress, an annealing experiment demonstrates the exclusive contribution of the dislocations for improved damage tolerance.

**Acknowledgement**

O.P. and J. R. thank for the financial support by DFG (No. 414179371). X. F. acknowledges the support by the Athene Young Investigator program at TU Darmstadt and the funding by the European Union (ERC, Project MECERDIS, Grant No. 101076167). Views and opinions expressed are, however, those of the authors only and do not necessarily reflect those of the European Union or the European Research Council. Neither the European Union nor the granting authority can be held responsible for them. W. Lu acknowledges the support by Shenzhen Science and Technology Program (grant no. JCYJ20230807093416034) and the use of the facilities at the Southern University of Science and Technology Core Research Facility. We thank Prof. K. Durst at TU Darmstadt for the access to the laser microscope and SEM, C. Okafor for the help with cross-sectional polishing and Prof. K. Higashida from Kyushu University for the very helpful discussions on crack tip-dislocation interaction. We would further like to thank P. Breckner for setting up the linear stage and writing a control software for the scratching method.

**Conflict of Interests:**

The authors declare no conflict of interests.